\documentclass[
reprint,
superscriptaddress,
 amsmath,amssymb,
 aps,
 pra,
]{revtex4-2}
\usepackage{graphicx}
\usepackage{overpic}
\usepackage{dcolumn}
\usepackage{bm}
\usepackage[hidelinks,colorlinks=true,linkcolor=blue,citecolor=blue]{hyperref}
\usepackage[mathlines]{lineno}
\usepackage{titlesec}
\usepackage{braket}
\usepackage{color}
\usepackage{float}
\usepackage[thicklines]{cancel}
\usepackage{soul}

\begin{document}


\title{In-situ Doppler-free spectroscopy with pulsed optical fields}

\author{Yuxin Wang$^{\dagger}$}
\affiliation{School of Physics and Beijing Key Laboratory of Opto-electronic Functional Materials and Micro-nano Devices, Renmin University of China, 100872 Beijing, China}

\author{Zhiyue Zheng$^{\dagger}$}
\affiliation{Beijing Academy of Quantum Information Sciences, 100193 Beijing, China}

\author{Qiuxin Zhang}
\affiliation{School of Physics and Beijing Key Laboratory of Opto-electronic Functional Materials and Micro-nano Devices, Renmin University of China, 100872 Beijing, China}

\author{Yonglang Lai}
\affiliation{School of Physics and Beijing Key Laboratory of Opto-electronic Functional Materials and Micro-nano Devices, Renmin University of China, 100872 Beijing, China}

\author{Zongqi Ge}
\affiliation{School of Physics and Beijing Key Laboratory of Opto-electronic Functional Materials and Micro-nano Devices, Renmin University of China, 100872 Beijing, China}

\author{Tianyi Wang}
\affiliation{School of Physics and Beijing Key Laboratory of Opto-electronic Functional Materials and Micro-nano Devices, Renmin University of China, 100872 Beijing, China}

\author{Liangyu Ding}
\affiliation{Beijing Academy of Quantum Information Sciences, 100193 Beijing, China}

\author{Smirnov Vasiliy}
\affiliation{P.N. Lebedev Physical Institute of the Russian Academy of Sciences, Moscow 119991, Russia}
\affiliation{Russian Quantum Center, Skolkovo, Moscow 121205, Russia}

\author{Ilya Semerikov}
\affiliation{P.N. Lebedev Physical Institute of the Russian Academy of Sciences, Moscow 119991, Russia}
\affiliation{Russian Quantum Center, Skolkovo, Moscow 121205, Russia}

\author{Shuaining Zhang}
\affiliation{School of Physics and Beijing Key Laboratory of Opto-electronic Functional Materials and Micro-nano Devices, Renmin University of China, 100872 Beijing, China}
\affiliation{Beijing Academy of Quantum Information Sciences, 100193 Beijing, China}
\affiliation{Key Laboratory of Quantum State Construction and Manipulation (Ministry of Education), Renmin University of China, 100872 Beijing, China}

\author{Wei Zhang}
\affiliation{School of Physics and Beijing Key Laboratory of Opto-electronic Functional Materials and Micro-nano Devices, Renmin University of China, 100872 Beijing, China}
\affiliation{Beijing Academy of Quantum Information Sciences, 100193 Beijing, China}
\affiliation{Key Laboratory of Quantum State Construction and Manipulation (Ministry of Education), Renmin University of China, 100872 Beijing, China}

\author{Xiang Zhang}
 \email{siang.zhang@ruc.edu.cn}
\affiliation{School of Physics and Beijing Key Laboratory of Opto-electronic Functional Materials and Micro-nano Devices, Renmin University of China, 100872 Beijing, China}
\affiliation{Beijing Academy of Quantum Information Sciences, 100193 Beijing, China}
\affiliation{Key Laboratory of Quantum State Construction and Manipulation (Ministry of Education), Renmin University of China, 100872 Beijing, China}

\date{\today}

\begin{abstract}

We propose a novel pulsed optical field method that alternately switches the pump beam in conventional saturation absorption to time-division multiplex the same probe beam into both probe and reference beams, followed by digital differential processing to achieve deterministic zero-background Doppler-free spectroscopy. This method effectively mitigates Doppler broadening and common-mode optical noise by addressing disturbances such as non-uniform background absorption and environmental noise, thereby offering enhanced accuracy and robustness. Using this technique, we measured the absolute frequency of Yb$^{+}$ isotopes in the $6s^2\ ^{1}S_0\to 6s6p ^{1}P_1$ transition. By employing an error signal derived from the first-derivative demodulated spectrum of $^{174}\mathrm{Yb}^{+}$, we achieved efficient stabilization of a 369.5 nm ultraviolet diode laser, demonstrating a frequency stability of $3 \times 10^{-11}$ over a 1500-second averaging period and a locking point uncertainty of 850 kHz sustained over 10 days. Furthermore, we report the first in-situ observation of Doppler-free Zeeman sub-level spectra, highlighting the precision of this method and its potential application in measuring magnetic field gradients.

\end{abstract}

\maketitle


\section{\label{sec:level1}Introduction}

Over the past few decades, a variety of methods have been developed in the fields of high-resolution spectroscopy and laser frequency stabilization. Among them, approaches based on Doppler-free atomic absorption have achieved significant advancements due to their simple experimental setups, combined with high absolute accuracy and rapid feedback capabilities. Typical approaches include saturated absorption spectroscopy\cite{PhysRevA.6.1280, Burd:14}, modulation transfer spectroscopy\cite{tanabe_frequency-stabilized_2018,wang_frequency_2011}, dichroic atomic vapor laser lock\cite{kim_frequency-stabilized_2003,sato_birefringent_2022}, and polarization spectroscopy\cite{Zhu14,lee-frequency-2014,10.1063/1.2973401}. In the conventional saturated absorption spectroscopy (SAS) scheme, the probe field overlapping with the pump field can only capture the Voigt profile \cite{BELAFHAL2000111}, therefore requiring the introduction of an additional reference field to obtain a Doppler-free spectrum. This approach exacerbates the system's complexity and makes it susceptible to inhomogeneous background absorption and environmental noise. Modulation transfer spectroscopy (MTS) reduces sensitivity to the Doppler background by applying high-frequency modulation to the pump beam and directly extracts the dispersion signal through two-beam interaction. However, the electro-optic modulator introduces unintended residual amplitude modulation (RAM) \cite{jaatinen_compensating_2008, negnevitsky_wideband_2013}, which can potentially distort the dispersion signal. The polarization spectroscopy (PS) method requires only two optical fields and does not involve modulation, it exclusively generates dispersion-like lineshapes, exhibits high sensitivity to the purity of laser polarization, and fails to effectively suppress common-mode noise.

To address these limitations, we present a more efficient and cost-effective technique: time-division multiplexing differential saturated absorption spectroscopy (TDMDSAS). This approach transforms spatially separated probe and reference beams into temporally distinct entities, enabling memory-based digital signal processing. A single photodetector effectively suppresses common-mode noise and enhances system robustness while enabling efficient spectrum measurement. Our method simultaneously captures high-resolution, noise-suppressed in-situ Doppler-free signals, together with demodulated signals for ultraviolet laser frequency stabilization. We measure the absolute frequency of the $^{1}S_0 \to ^{1}$$P$$_1$ transition in $\mathrm{Yb}^{+}$ isotopes, and perform frequency stabilization on the 369.5 nm ultraviolet laser. The ultraviolet laser is locked to the transition of $^{174}\mathrm{Yb}^{+}$ at 811.2915891 THz, achieving stability of 26 kHz with an averaging time of 1500 s, and a locking point uncertainty of 850 kHz over 10 days. Additionally, we achieved the first in-situ observation of Doppler-free Zeeman sub-level spectra, facilitating the precise measurement of magnetic field gradients. Compared with the single-beam SAS method\cite{Wang2022,sargsyan_approach_2019}, which improves traditional SAS by employing a space-division multiplexing approach, our method is immune to challenges related to structural stability and common-mode noise due to the reliance on two photodetectors (PD). Our innovative method holds immense potential for extension to other atomic species and offers practical significance for applications in atomic clocks\cite{mcgrew_atomic_2018}, atomic magnetometers\cite{ma-situ_2023} and quantum simulation\cite{PhysRevLett.126.083604}.

\begin{figure}
	\includegraphics[width=0.46\textwidth]{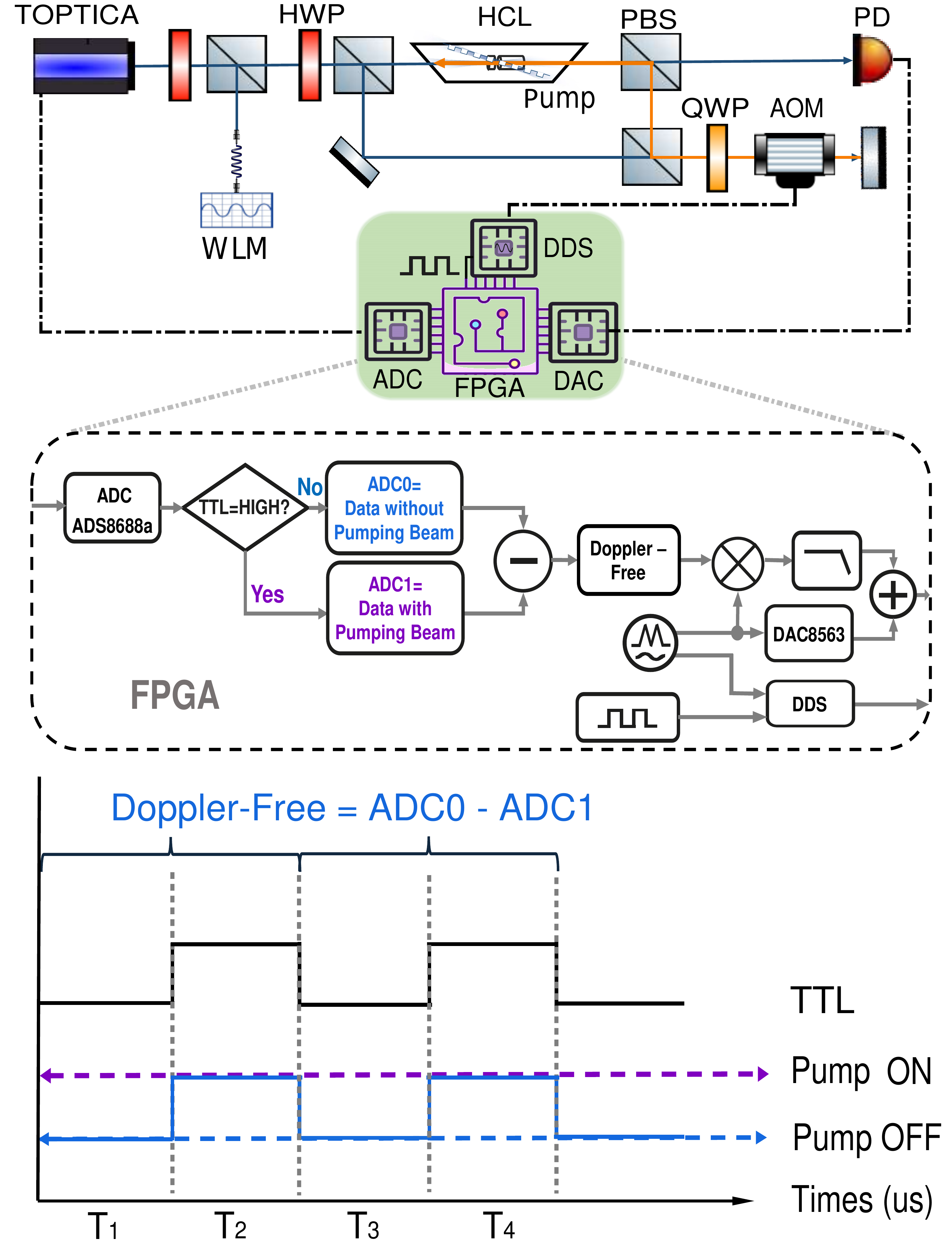}\\[5pt]  
	\caption{(Color online) Schematic diagram of the experimental setup: half-wave plate (HWP), polarization beam splitter (PBS), hollow cathode lamp (HCL), acousto-optic modulator (AOM), photodetector (PD), wavelength meter(WS-8). The dashed box illustrates the process of internal signal digitized processing within the FPGA. The bottom figure illustrates the timing sequence used in the signal processing.}
	\label{fig1}
\end{figure}

\section{EXPERIMENTAL}

In the TDMDSAS setup, the pump beam is pulse amplitude modulated. A low TTL signal indicates the absence of the pump beam, causing a pronounced broadening of the atomic absorption spectrum attributed to the Doppler effect. This broadening results in the Doppler broadening profile, as outlined in Eq.\ref{eq1}.
\begin{equation} 
	\label{eq1}
	D(\omega,\omega_{0})=\frac{2 \sqrt{\ln 2}}{\sqrt{\pi}\Gamma}\exp \left[\frac{-2\sqrt{\ln 2} (\omega-\omega_{0})^{2}}{\Gamma}\right]
\end{equation}
A high TTL signal indicates the presence of the pump laser, where different velocity groups resonate with the beams. The higher-power pump beam induces saturated absorption fatigue in the probe field, resulting in the characteristic SAS profile as described in Eq.\ref{eq2}.  
\begin{equation} 
	\label{eq2}
		S(\omega)=\frac{\gamma/2\pi}{(\omega-\omega_{0})^{2}+(\gamma/2)^{2}}+D(\omega,\omega_{0})
\end{equation}
where $\omega$ is the laser frequency and $\omega_{0}$ is the atomic transition frequency. $\gamma$ and $\Gamma$ represent the full width at half maximum (FWHM) of the Lorentzian and Gaussian line shapes, respectively. The received signal undergoes real-time subtraction to produce the Doppler-free signal 
 $S(\omega)-D(\omega)$ as the TTL signal alternates between its states.

The experimental setup is illustrated in Fig.~\ref{fig1}. A 369.5 nm ultraviolet diode laser (TOPTICA DLC Pro) is directed through two polarization beam splitters (PBS), splitting its output into three distinct beams. One beam is directed to the wavelength meter (HighFinesse WS-8, WLM) for frequency measurement, while the remaining two beams serve as the probe and pump beams. Both beams are tightly focused through a lens into a hollow cathode lamp (Hamamatsu L2783, HCL). To achieve a Doppler-free spectrum with an optimal signal-to-noise ratio, we finely tuned the power and polarization using PBS and half-wave plates (HWP) and optimized the operating current of the Yb HCL. Upon finalizing these parameters, the HCL's driving voltage was set to 197 V with a stable current of 15 mA. Directly before entering the HCL, the intensity of the probe beam was maintained at 600 $\mu\mathrm{W}$, and that of the pump beam at 1 mW. Within the atomic sample, the waist radius of both the probe and pump beams was approximately 1 $\mathrm{mm}$. Precise alignment of the probe beam with the counter-propagating pump beam is critical to avoid RAM\cite{jaatinen_compensating_2008,jaatinen_residual_2009,sun_modulation_2016,shen_systematic_2015}, which could distort the signal. After passing through the HCL, the probe beam is detected by a photodetector (PDA100A, PD) with a 50 dB gain. The acquired signal is then processed by an analog-to-digital converter (ADC) and subsequently analyzed by an FPGA\cite{carminati_impact_2021,avalos_field-programmable-gate-array-based_2023}.

The double-pass configuration of the pump beam through the acousto-optic modulator (SGT200-397-1TA, AOM) \cite{wu_modulation_2018}, utilizing the $+1$ and $-1$ diffraction orders, results in a zero frequency shift for the pump beam. This configuration ensures that the spectroscopic features are centered at the bare transition frequencies. The AOM is controlled by a TTL signal from the FPGA via a direct digital synthesis (DDS). The high and low states of the TTL signal determine the presence or absence of the pump beam. We selected a TTL signal frequency of 20 kHz, taking into account the response time of the atoms, PD, AOM, and potential interleaving artifacts in signal processing. The FPGA performs time-domain differentiation based on the state of the TTL signal to extract Doppler-free signals. The two difference signals originate from the same probe beam, enabling the acquisition of in-situ Doppler background information within the pump-probe region for the same atom ensemble. Additionally, by employing time-interleaved sampling of the same signal while maintaining a shared sampling rate, high-speed data acquisition is achieved without compromising signal integrity.

\begin{figure}
	\includegraphics[width=0.5\textwidth]{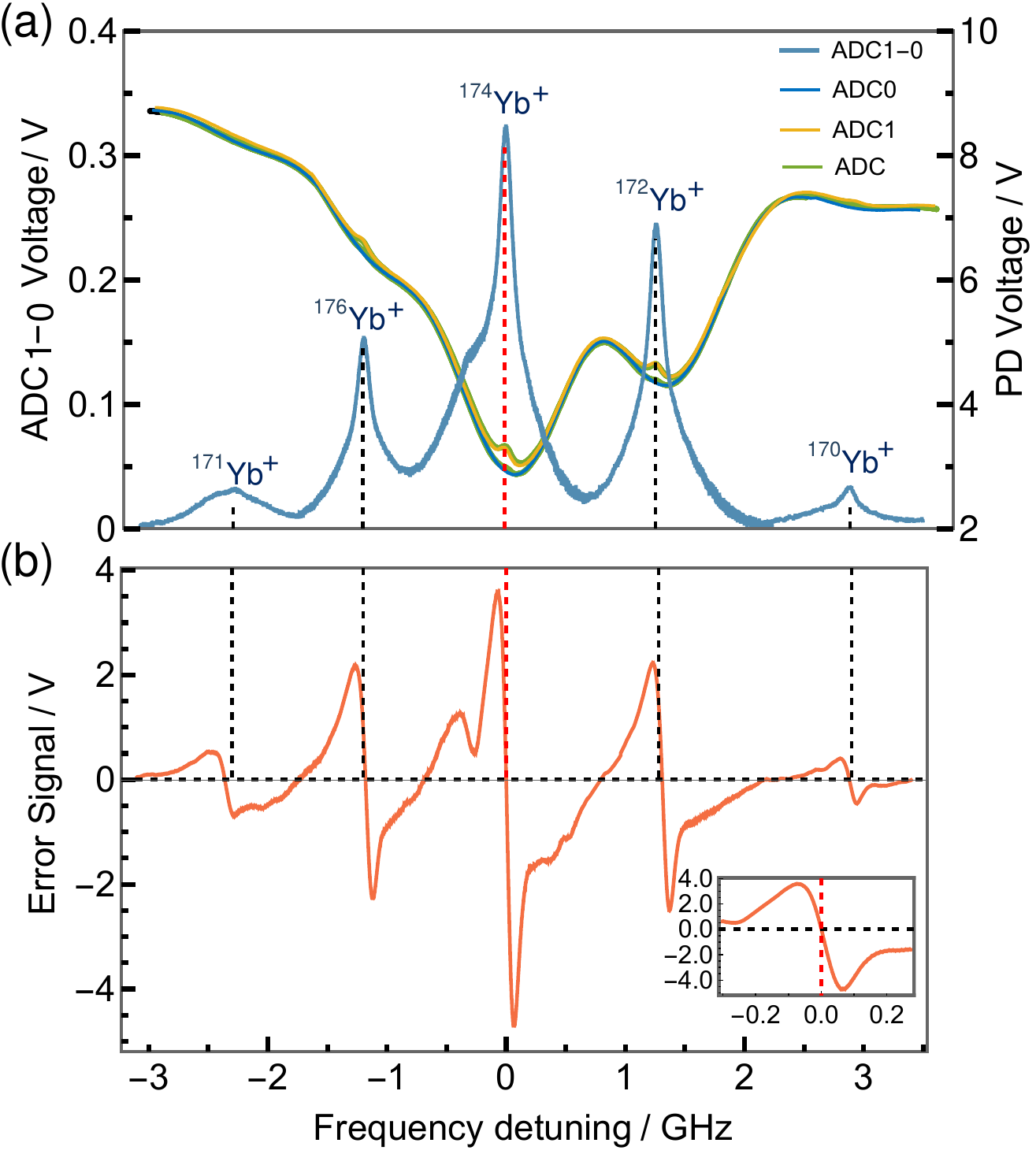}\\[5pt]  
	\caption{(Color online) (a) The spectrum of $^{1}S_{0} \to ^{1}P_{1}$ transition of Yb, $\mathrm{V}_{\mathrm{ADC}}$(green) represents all the signals received from the PD, $\mathrm{V}_\mathrm{ADC1}$ (yellow), indicative of Doppler- broadened absorption while the TTL is in a low state, $\mathrm{V}_{\mathrm{ADC0}}$(blue) corresponds to saturated absorption peaks with a high TTL signal, and the Doppler-free spectrum $\mathrm{V}_{\mathrm{ADC1-0}}$ (light blue) is obtained by differencing $\mathrm{V}_{\mathrm{ADC0}}$ from $\mathrm{V}_{\mathrm{ADC1}}$. (b) The associated first-derivative demodulation signal is derived from $\mathrm{V}_{\mathrm{ADC1-0}}$. The inset shows a detailed image of the demodulation signal of $^{174}\mathrm{Yb}^{+}$ for laser frequency stabilization. The detunings are relative to the absolute frequency of the $^{174}\mathrm{Yb}^{+}$ transition.}
	\label{fig2}
\end{figure}

To acquire the error signal sent to the feedback loop for frequency stabilization, we apply a modulation signal, $A\sin(\Omega t)$, and set an operating point for demodulating the signal with the first harmonic, which can be described by
\begin{equation}
	\begin{aligned}
		& \frac{1}{T}\int_{0}^{T}\sin(\Omega t) S \left[ \omega_{0}+ A \sin(\Omega t)\right] \\
		& = \left. \frac{A}{2}\left\{\frac{-\gamma(\omega-\omega_{0})}{\pi \left[(\omega-\omega_{0})^{2}+(\gamma/2)^{2}\right]^{2}}  + D^{(1)}(\omega,\omega_{0})\right\}\right|_{\omega}
	\end{aligned}
	\label{eq3}
\end{equation}
where $T=2\pi/\Omega$ and $D^{(1)}(\omega,\omega_{0})$ is the first derivative of $D(\omega,\omega_{0})$, and $\mathrm{A}$ is the modulation depth. During the demodulation process, we scan $\omega$ to generate a dispersive-like lineshape. To achieve a superior Doppler-free spectrum, we aim to minimize the Doppler background as effectively as possible. Experimentally, we apply a 47 kHz sine wave modulation signal to the laser's diode voltage to generate the error signal for laser frequency stabilization. We derive the error signal by mixing the Doppler-free signal with the modulation signal and processing it through a digital lock-in amplifier. The error signal is then analyzed by the proportional-integral-differential (PID) controller and fed back to the laser controller for frequency stabilization.

Specifically, the FPGA categorizes the received signal, $\mathrm{V}_{\mathrm{ADC}}$, from the PD into three distinct types based on the TTL signal's states. As indicated in Fig.~\ref{fig2} (a), these include the Doppler-broadened absorption signal $\mathrm{V}_{\mathrm{ADC0}}$, the saturated absorption signal $\mathrm{V}_{\mathrm{ADC1}}$, and the Doppler-free signal $\mathrm{V}_{\mathrm{ADC1-0}}$. Furthermore, Fig.~\ref{fig2}(b) displays the spectral first-derivative demodulated spectrum of the $\mathrm{V}_{\mathrm{ADC1-0}}$, as explained in the first term on the right side of Eq.\ref{eq3}. The controller board, incorporating FPGA (Xilinx Spartan6 LX9), DAC (DAC8563), and ADC(ADS8688A), integrates multiple functions including time-domain signal processing, digital lock-in amplifier, digital PID controller, digital filter, and digital signal generation. Moreover, the controller board and highly engineered interactive software play a crucial role in acquiring Doppler-free spectra and stabilizing laser frequency.

\section{RESULTS}

\subsection{Doppler-free spectroscopy}

\begin{table}
	\centering
	\caption{Absolute transition frequencies of the $^{1}S_{0} \to ^{1}$$P$$_{1}$ transition lines}
	\begin{ruledtabular}
		\begin{tabular}{ccc}
			Isotope & Frequency (THz) \\
			\colrule
			\\
			{$^{171}\mathrm{Yb}^{+}(1-0)$}           & {811.289268526(3)}    \\
			{$^{176}\mathrm{Yb}^{+}$}               & {811.290375053(3)}    \\
			{$^{174}\mathrm{Yb}^{+}$}               & {811.291589162(3)}    \\
			{$^{172}\mathrm{Yb}^{+}$}               & {811.292869140(3)}     \\
			{$^{170}\mathrm{Yb}^{+}$}               & {811.294493818(3)}    \\
		\end{tabular}
	\end{ruledtabular}
    \label{tab1}
\end{table}

\begin{figure}
	\hspace{-2em}
	\includegraphics[width=0.5\textwidth]{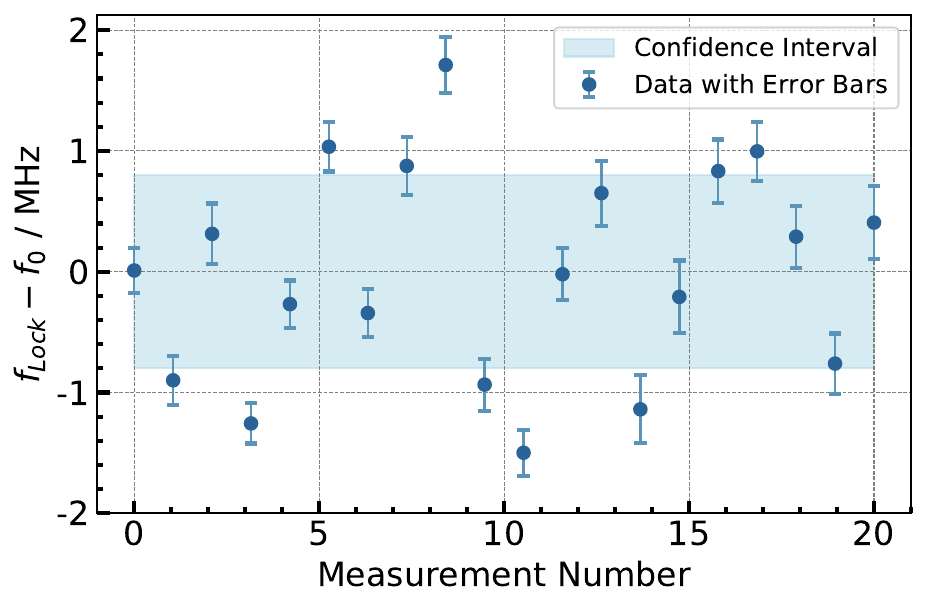}\\[5pt]  
	\caption{(Color online) Daily variations in the lock point are shown for the laser locked to the $^{174}\mathrm{Yb}^{+}$ transition using TDMDSAS. Each frequency data point represents an average of 300 measurements taken by the WLM. The individual error bars represent the standard deviation of the respective measurement set.}
	\label{fig3}
\end{figure}

Resonance frequencies measured with our method for different isotopes of Yb are presented in Table~\ref{tab1}. The absolute frequency of $^{174}\mathrm{Yb}^{+}$ is determined to be $811.2915891(3)$ THz. The absolute frequency uncertainty (0.3 MHz) is calculated from 10 measurements conducted over one day. Fig.~\ref{fig3} shows 20 frequency measurements recorded by the WLM over a 10-day period. Each data point contains a 2-minute frequency recording on the WLM with a 30~ms exposure time. Notably, the experimental system remained unaltered throughout this period, and the statistical uncertainty stayed within 850~kHz. Compared to Ref. \cite{lee-frequency-2014}, the absolute frequency shift is approximately 100~MHz. This shift is primarily due to the WLM's absolute uncertainty, which is sensitive to environmental temperature fluctuations and air pressure disturbances\cite{PhysRevA.60.1103, PhysRevA.101.062506}. It is also potentially attributed to laser power and the operating parameters of the HCL\cite{PhysRevA.89.062510, qiao_investigation_2023}.

\subsection{Laser frequency stabilization}

\begin{figure}
	\includegraphics[width=0.5\textwidth]{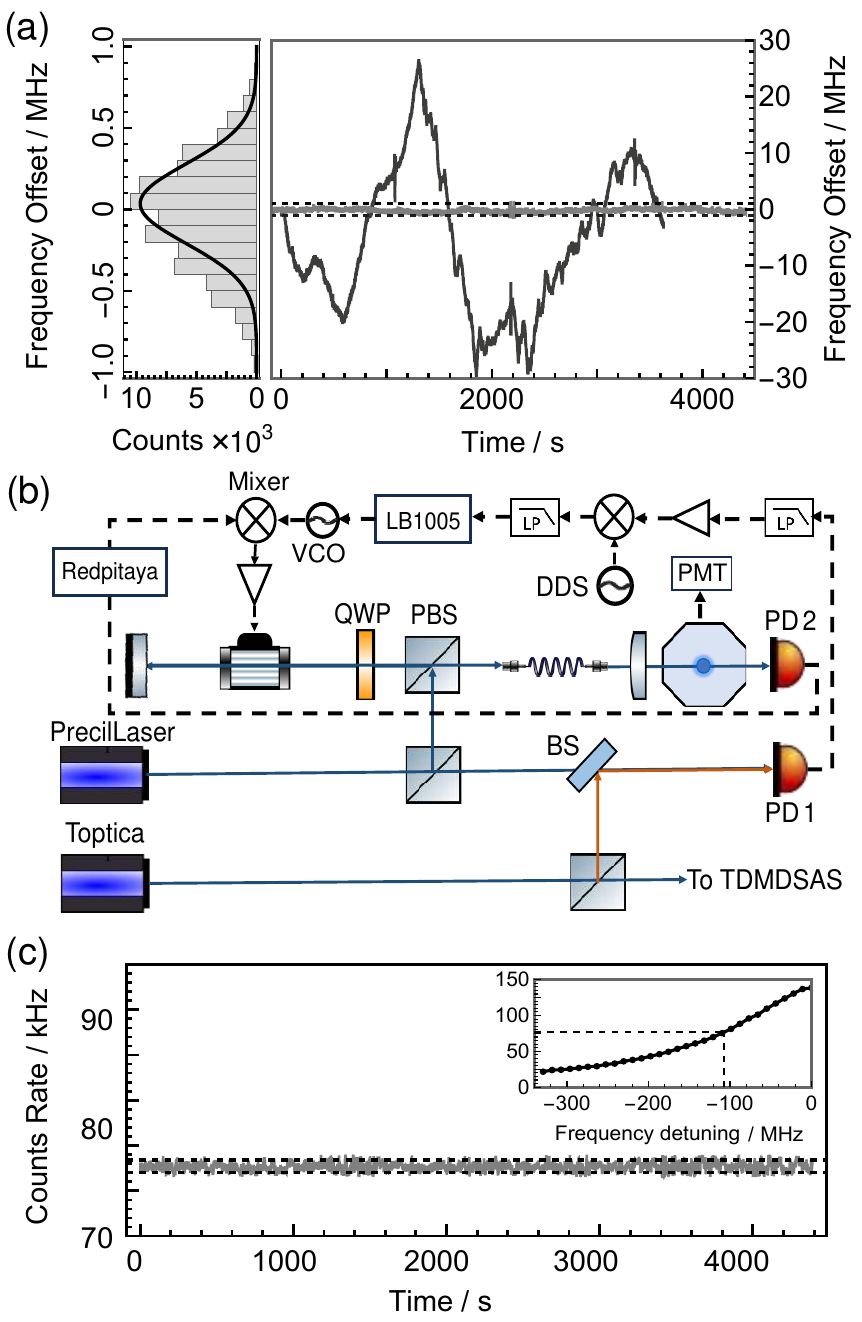}\\[5pt]  
	\caption{(Color online) (a) Frequency drift of the laser when free-running (black) and locked to the $^{174}\mathrm{Yb}^{+}$ transition using TDMDSAS (gray). A Gaussian histogram illustrates frequency offsets from the central frequencies when the laser is locked. (b) The schematic diagram for obtaining the ion fluorescence signal includes cooling beam power stabilization, the beat frequency method for locking the cooling beam frequency, and the ion trap system. In this setup, PD1 is a photodetector with a fixed gain (PDA10A2). The cooling beam passes through the double-pass configuration, and its power and frequency are reactively adjusted by the AOM in real time. PMT: photomultiplier, BS: beamsplitters, Amp: amplifier, VCO: voltage controlled oscillator. (c) The ion fluorescence count rate (black) remains stable. The inset shows the fluorescence spectrum as a function of frequency detuning in MHz.}
	\label{fig4}
\end{figure}

To quantify the frequency stability and achieve optimal laser frequency stabilization, we locked the laser at the zero-crossing point of the dispersion signal corresponding to the transitions of $^{174}\mathrm{Yb}^{+}$ ($^{1}S_{0} \to ^{1}$$P$$_{1}$) \cite{PhysRevA.82.063419}, as shown in the inset of Fig.~\ref{fig2} (b). This process stabilized the target laser at a wavelength of 369.52491805~nm over an extended period. In Fig.~\ref{fig4} (a), black points represent the drift measurements of the laser in a free-running state, while gray points represent the drift measurements of the laser when locked using TDMDSAS, revealing a frequency standard deviation of 338~kHz over 4000~s. The histogram on the left side illustrates the deviations in the laser frequency from its central frequency, with an FWHM of 0.79~MHz.

We further utilize the Doppler cooling scattering fluorescence response of a single $^{174}\mathrm{Yb}^{+}$ ion trapped in the blade trap\cite{blade1,blade2} as a frequency discriminator to evaluate the stability of TDMDSAS, as shown in Fig.~\ref{fig4} (b). We transmit the frequency of the locked ion cooling beam using the beat frequency method \cite{4785283, Uehara2014OpticalBF}, and the frequency difference between the transition frequency of $^{174}\mathrm{Yb}^{+}$ in the cathode lamp and the cooling beam incident on the trap is approximately 105~MHz. The cooling beam power is approximately 50~$\mu$W, and the fluorescence scattered by the ions is detected by a photomultiplier tube (PMT), with each measurement lasting 1000~$\mu$s. Fig.~\ref{fig4} (c) shows the variation of ion fluorescence over a 4000~s interval, with each data point representing the average of 1000 measurements.

\begin{figure}
	\includegraphics[width=0.5\textwidth]{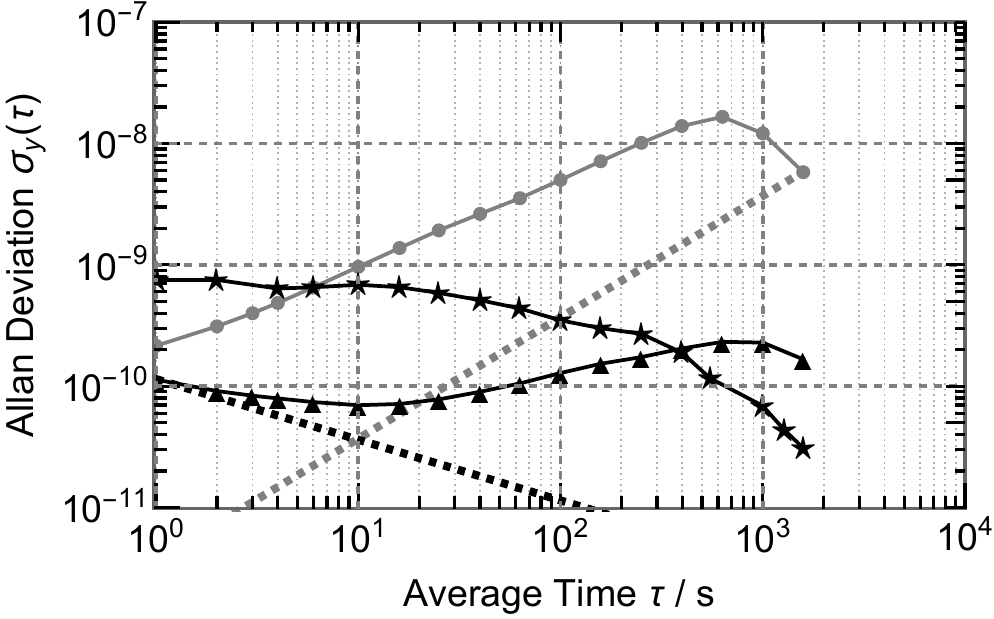}\\[5pt]  
	\caption{(Color online) Allan deviation $\sigma_{y} (\tau)$ of the laser frequency fluctuations for free-running operation (gray circles) and locked to the $\sigma_{y} (\tau)$ transition in $^{174}\mathrm{Yb}^{+}$ using TDMDSAS, with measurements recorded by the WLM (black triangles) and calibrated based on fluorescence count rates (black stars). The gray dashed line represents long-term frequency drift, while the black dashed line denotes white noise in frequency stabilization.}
	\label{fig5}
\end{figure}

We are interested in calculating the Allan deviation $\sigma_{y} (\tau)$ of the frequency fluctuation for free-running operation and frequency stabilization measured by the WLM and the fluorescence count rate of a single trapped ion, as shown in Fig.~\ref{fig5}. Regarding the free-running drift, the Allan deviation follows a $\tau^{1}$ dependence. When the laser is locked using TDMDSAS, for the WLM measurement, the Allan deviation was initially 93~kHz at $\tau = 1$~s, decreased to 56~kHz at $\tau = 100$~s, but eventually increased to 185~kHz at $\tau = 1000$~s, as shown in Fig.~\ref{fig5}. This abnormal behavior could arise from long-term drifts in the WLM itself \cite{Shen_2020, Saleh:15, 10.1063/1.5025537}. For the fluorescence count rate of a single trapped ion, the obtained Allan deviation is 610~kHz ($7.5 \times 10^{-11}$) at an averaging time of $\tau = 1$~s, and it generally follows the $\tau^{-1/2}$ slope from $\tau =10$~s. A frequency stability of 26~kHz ($3.0 \times 10^{-11}$) is achieved at an averaging time of 1500~s.

\begin{figure}
	\includegraphics[width=0.5\textwidth]{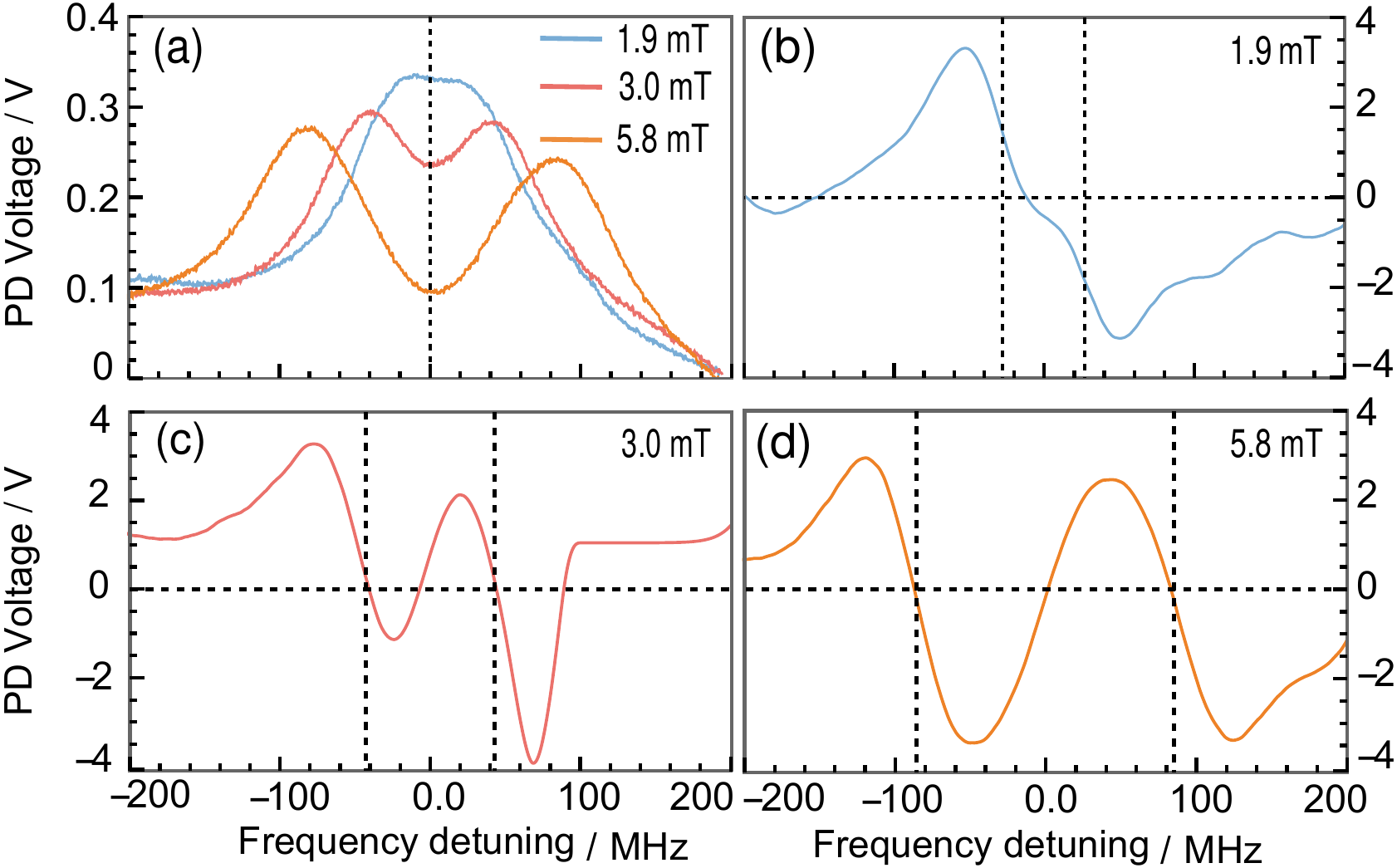}\\[5pt]  
	\caption{(Color online) (a) TDMDSAS spectra recorded at magnetic field strengths of 1.9, 3.0, and 5.8 mT. (b), (c), and (d) show the first-derivative demodulated signals corresponding to the three magnetic field strengths. Detunings are measured relative to the absolute frequency of the $^{174}\mathrm{Yb}^{+}$ transition.}
	\label{fig6}
\end{figure}

\subsection{Zeeman sub-level spectroscopy}

The linewidth of the spectroscopy for the $^{1}S_{0} \to ^{1}$$P$$_{1}$ transition in $^{174}\mathrm{Yb}^{+}$ measured by TDMDSAS is approximately 80~MHz, broader than the natural linewidth of 20~MHz, reaching the minimum SAS resolution\cite{tanabe_frequency-stabilized_2018}. The broadening observed in the Doppler-free signal is attributed to collisional dephasing induced by the buffer gas (Ne) sealed in the HCL\cite{Moon:18, Heung-Ryoul}. To further ensure the minimum resolution of TDMDSAS, we applied external magnetic fields aligned parallel to the probe beam using a permanent magnet. This setup allowed the observation of Doppler-free spectra featuring Zeeman sub-levels and the absolute Zeeman splitting frequency more accurately. The magnetic field is adjusted by altering the placement of the permanent magnet, and it is important to note that the magnetic shielding effect of the hollow cathode should be considered \cite{sato_birefringent_2022,kim_frequency-stabilized_2003}.

As illustrated in Fig.~\ref{fig6}, we investigated the Doppler-free spectroscopy under diverse magnetic field strengths of 1.9, 3.0, and 5.8~mT, with corresponding Zeeman sub-level shifts of 27, 43, and 82~MHz. As observedd in Fig.~\ref{fig6} (b), when the magnetic field reaches 1.9~mT, the Zeeman splitting at a frequency of 27~MHz remains distinguishable. The resolution has reached its optimum limits due to the constraints imposed by the natural linewidth of the ion. Our method, engineered for precision measurement of both the magnetic field and its gradient, effectively navigates around the hurdles inherent in conventional SAS techniques, where magnetic field disparities arise due to the spatial separation between the probe and reference beams.

\section{CONCLUSION}

In summary, a straightforward, versatile, and efficient TDMDSAS method has been innovatively established. This method effectively capture in-situ Doppler background information, including direction, this scheme exhibits excellent noise suppression on the Doppler-free spectrum with zero background latching. This capability contributes to Doppler-free spectra with exceptional long-term absolute stability and enhanced resolution. Our efforts encompass absolute frequency measurements of Yb$^{+}$ isotopes, exploration of Doppler-free spectra with Zeeman levels, and the achievement of high-performance laser frequency stabilization. Furthermore, the in-situ magnetic field measurement functionality broadens its application potential for magnetic field gradient determination. This protocol is widely applicable to laser cooling experiments and can be extended to various types of lasers and atomic species. It presents significant implications for quantum computation with atoms and ions, as well as practical value in applications such as atomic clocks and atomic magnetometers.


\section*{ACKNOWLEDGMENTS}
The authors acknowledge financial support from the National Natural Science Foundation of China (12074427, 12074428,  12204535, 12304565, 92265208); National Key R\&D Program of China (2022YFA1405300).



\nocite{*}
\bibliographystyle{apsrev4-2}
\bibliography{references.bib}

\begin{thebibliography}{43}%
\makeatletter
\providecommand \@ifxundefined [1]{%
 \@ifx{#1\undefined}
}%
\providecommand \@ifnum [1]{%
 \ifnum #1\expandafter \@firstoftwo
 \else \expandafter \@secondoftwo
 \fi
}%
\providecommand \@ifx [1]{%
 \ifx #1\expandafter \@firstoftwo
 \else \expandafter \@secondoftwo
 \fi
}%
\providecommand \natexlab [1]{#1}%
\providecommand \enquote  [1]{``#1''}%
\providecommand \bibnamefont  [1]{#1}%
\providecommand \bibfnamefont [1]{#1}%
\providecommand \citenamefont [1]{#1}%
\providecommand \href@noop [0]{\@secondoftwo}%
\providecommand \href [0]{\begingroup \@sanitize@url \@href}%
\providecommand \@href[1]{\@@startlink{#1}\@@href}%
\providecommand \@@href[1]{\endgroup#1\@@endlink}%
\providecommand \@sanitize@url [0]{\catcode `\\12\catcode `\$12\catcode
  `\&12\catcode `\#12\catcode `\^12\catcode `\_12\catcode `\%12\relax}%
\providecommand \@@startlink[1]{}%
\providecommand \@@endlink[0]{}%
\providecommand \url  [0]{\begingroup\@sanitize@url \@url }%
\providecommand \@url [1]{\endgroup\@href {#1}{\urlprefix }}%
\providecommand \urlprefix  [0]{URL }%
\providecommand \Eprint [0]{\href }%
\providecommand \doibase [0]{https://doi.org/}%
\providecommand \selectlanguage [0]{\@gobble}%
\providecommand \bibinfo  [0]{\@secondoftwo}%
\providecommand \bibfield  [0]{\@secondoftwo}%
\providecommand \translation [1]{[#1]}%
\providecommand \BibitemOpen [0]{}%
\providecommand \bibitemStop [0]{}%
\providecommand \bibitemNoStop [0]{.\EOS\space}%
\providecommand \EOS [0]{\spacefactor3000\relax}%
\providecommand \BibitemShut  [1]{\csname bibitem#1\endcsname}%
\let\auto@bib@innerbib\@empty
\bibitem [{\citenamefont {Haroche}\ and\ \citenamefont
  {Hartmann}(1972)}]{PhysRevA.6.1280}%
  \BibitemOpen
  \bibfield  {author} {\bibinfo {author} {\bibfnamefont {S.}~\bibnamefont
  {Haroche}}\ and\ \bibinfo {author} {\bibfnamefont {F.}~\bibnamefont
  {Hartmann}},\ }\href {https://doi.org/10.1103/PhysRevA.6.1280} {\bibfield
  {journal} {\bibinfo  {journal} {Phys. Rev. A}\ }\textbf {\bibinfo {volume}
  {6}},\ \bibinfo {pages} {1280} (\bibinfo {year} {1972})}\BibitemShut
  {NoStop}%
\bibitem [{\citenamefont {Burd}\ \emph {et~al.}(2013)\citenamefont {Burd},
  \citenamefont {du~Toit},\ and\ \citenamefont {Uys}}]{Burd:14}%
  \BibitemOpen
  \bibfield  {author} {\bibinfo {author} {\bibfnamefont {S.~C.}\ \bibnamefont
  {Burd}}, \bibinfo {author} {\bibfnamefont {P.~J.~W.}\ \bibnamefont
  {du~Toit}},\ and\ \bibinfo {author} {\bibfnamefont {H.}~\bibnamefont {Uys}},\
  }\href {https://api.semanticscholar.org/CorpusID:10028409} {\bibfield
  {journal} {\bibinfo  {journal} {Optics express}\ }\textbf {\bibinfo {volume}
  {22 21}},\ \bibinfo {pages} {25043} (\bibinfo {year} {2013})}\BibitemShut
  {NoStop}%
\bibitem [{\citenamefont {Tanabe}\ \emph {et~al.}(2018)\citenamefont {Tanabe},
  \citenamefont {Akamatsu}, \citenamefont {Inaba}, \citenamefont {Okubo},
  \citenamefont {Kobayashi}, \citenamefont {Yasuda}, \citenamefont {Hosaka},\
  and\ \citenamefont {Hong}}]{tanabe_frequency-stabilized_2018}%
  \BibitemOpen
  \bibfield  {author} {\bibinfo {author} {\bibfnamefont {T.}~\bibnamefont
  {Tanabe}}, \bibinfo {author} {\bibfnamefont {D.}~\bibnamefont {Akamatsu}},
  \bibinfo {author} {\bibfnamefont {H.}~\bibnamefont {Inaba}}, \bibinfo
  {author} {\bibfnamefont {S.}~\bibnamefont {Okubo}}, \bibinfo {author}
  {\bibfnamefont {T.}~\bibnamefont {Kobayashi}}, \bibinfo {author}
  {\bibfnamefont {M.}~\bibnamefont {Yasuda}}, \bibinfo {author} {\bibfnamefont
  {K.}~\bibnamefont {Hosaka}},\ and\ \bibinfo {author} {\bibfnamefont {F.-L.}\
  \bibnamefont {Hong}},\ }\href {https://doi.org/10.7567/JJAP.57.062501}
  {\bibfield  {journal} {\bibinfo  {journal} {Japanese Journal of Applied
  Physics}\ }\textbf {\bibinfo {volume} {57}},\ \bibinfo {pages} {062501}
  (\bibinfo {year} {2018})}\BibitemShut {NoStop}%
\bibitem [{\citenamefont {and}\ \emph {et~al.}(2011)\citenamefont {and}, , ,\
  and\ \citenamefont {and}}]{wang_frequency_2011}%
  \BibitemOpen
  \bibfield  {author} {\bibinfo {author} {\bibnamefont {and}}, , ,\ and\
  \bibinfo {author} {\bibnamefont {and}},\ }\href
  {https://doi.org/10.1088/1674-1056/20/1/013201} {\bibfield  {journal}
  {\bibinfo  {journal} {Chinese Physics B}\ }\textbf {\bibinfo {volume} {20}},\
  \bibinfo {pages} {013201} (\bibinfo {year} {2011})}\BibitemShut {NoStop}%
\bibitem [{\citenamefont {Kim}\ \emph {et~al.}(2003)\citenamefont {Kim},
  \citenamefont {Park}, \citenamefont {Yeom}, \citenamefont {Kim},\ and\
  \citenamefont {Yoon}}]{kim_frequency-stabilized_2003}%
  \BibitemOpen
  \bibfield  {author} {\bibinfo {author} {\bibfnamefont {J.~I.}\ \bibnamefont
  {Kim}}, \bibinfo {author} {\bibfnamefont {C.~Y.}\ \bibnamefont {Park}},
  \bibinfo {author} {\bibfnamefont {J.~Y.}\ \bibnamefont {Yeom}}, \bibinfo
  {author} {\bibfnamefont {E.~B.}\ \bibnamefont {Kim}},\ and\ \bibinfo {author}
  {\bibfnamefont {T.~H.}\ \bibnamefont {Yoon}},\ }\href
  {https://doi.org/10.1364/OL.28.000245} {\bibfield  {journal} {\bibinfo
  {journal} {Optics Letters}\ }\textbf {\bibinfo {volume} {28}},\ \bibinfo
  {pages} {245} (\bibinfo {year} {2003})}\BibitemShut {NoStop}%
\bibitem [{\citenamefont {Sato}\ \emph {et~al.}(2022)\citenamefont {Sato},
  \citenamefont {Hayakawa}, \citenamefont {Okamoto}, \citenamefont {Shimomura},
  \citenamefont {Aoki},\ and\ \citenamefont {Torii}}]{sato_birefringent_2022}%
  \BibitemOpen
  \bibfield  {author} {\bibinfo {author} {\bibfnamefont {T.}~\bibnamefont
  {Sato}}, \bibinfo {author} {\bibfnamefont {Y.}~\bibnamefont {Hayakawa}},
  \bibinfo {author} {\bibfnamefont {N.}~\bibnamefont {Okamoto}}, \bibinfo
  {author} {\bibfnamefont {Y.}~\bibnamefont {Shimomura}}, \bibinfo {author}
  {\bibfnamefont {T.}~\bibnamefont {Aoki}},\ and\ \bibinfo {author}
  {\bibfnamefont {Y.}~\bibnamefont {Torii}},\ }\href
  {https://doi.org/10.1364/JOSAB.442465} {\bibfield  {journal} {\bibinfo
  {journal} {JOSA B}\ }\textbf {\bibinfo {volume} {39}},\ \bibinfo {pages}
  {155} (\bibinfo {year} {2022})}\BibitemShut {NoStop}%
\bibitem [{\citenamefont {Zhu}\ \emph {et~al.}(2014)\citenamefont {Zhu},
  \citenamefont {Chen}, \citenamefont {Li},\ and\ \citenamefont
  {Wang}}]{Zhu14}%
  \BibitemOpen
  \bibfield  {author} {\bibinfo {author} {\bibfnamefont {S.}~\bibnamefont
  {Zhu}}, \bibinfo {author} {\bibfnamefont {T.}~\bibnamefont {Chen}}, \bibinfo
  {author} {\bibfnamefont {X.}~\bibnamefont {Li}},\ and\ \bibinfo {author}
  {\bibfnamefont {Y.}~\bibnamefont {Wang}},\ }\href
  {https://doi.org/10.1364/JOSAB.31.002302} {\bibfield  {journal} {\bibinfo
  {journal} {J. Opt. Soc. Am. B}\ }\textbf {\bibinfo {volume} {31}},\ \bibinfo
  {pages} {2302} (\bibinfo {year} {2014})}\BibitemShut {NoStop}%
\bibitem [{\citenamefont {Lee}\ \emph {et~al.}(2014)\citenamefont {Lee},
  \citenamefont {Jarratt}, \citenamefont {Marciniak},\ and\ \citenamefont
  {Biercuk}}]{lee-frequency-2014}%
  \BibitemOpen
  \bibfield  {author} {\bibinfo {author} {\bibfnamefont {M.~W.}\ \bibnamefont
  {Lee}}, \bibinfo {author} {\bibfnamefont {M.~C.}\ \bibnamefont {Jarratt}},
  \bibinfo {author} {\bibfnamefont {C.}~\bibnamefont {Marciniak}},\ and\
  \bibinfo {author} {\bibfnamefont {M.~J.}\ \bibnamefont {Biercuk}},\ }\href
  {https://doi.org/10.1364/OE.22.007210} {\bibfield  {journal} {\bibinfo
  {journal} {Optics Express}\ }\textbf {\bibinfo {volume} {22}},\ \bibinfo
  {pages} {7210} (\bibinfo {year} {2014})}\BibitemShut {NoStop}%
\bibitem [{\citenamefont {Streed}\ \emph {et~al.}(2008)\citenamefont {Streed},
  \citenamefont {Weinhold},\ and\ \citenamefont
  {Kielpinski}}]{10.1063/1.2973401}%
  \BibitemOpen
  \bibfield  {author} {\bibinfo {author} {\bibfnamefont {E.~W.}\ \bibnamefont
  {Streed}}, \bibinfo {author} {\bibfnamefont {T.~J.}\ \bibnamefont
  {Weinhold}},\ and\ \bibinfo {author} {\bibfnamefont {D.}~\bibnamefont
  {Kielpinski}},\ }\href {https://doi.org/10.1063/1.2973401} {\bibfield
  {journal} {\bibinfo  {journal} {Applied Physics Letters}\ }\textbf {\bibinfo
  {volume} {93}},\ \bibinfo {pages} {071103} (\bibinfo {year}
  {2008})}\BibitemShut {NoStop}%
\bibitem [{\citenamefont {Belafhal}(2000)}]{BELAFHAL2000111}%
  \BibitemOpen
  \bibfield  {author} {\bibinfo {author} {\bibfnamefont {A.}~\bibnamefont
  {Belafhal}},\ }\href
  {https://doi.org/https://doi.org/10.1016/S0030-4018(00)00564-2} {\bibfield
  {journal} {\bibinfo  {journal} {Optics Communications}\ }\textbf {\bibinfo
  {volume} {177}},\ \bibinfo {pages} {111} (\bibinfo {year}
  {2000})}\BibitemShut {NoStop}%
\bibitem [{\citenamefont {Jaatinen}\ and\ \citenamefont
  {Hopper}(2008)}]{jaatinen_compensating_2008}%
  \BibitemOpen
  \bibfield  {author} {\bibinfo {author} {\bibfnamefont {E.}~\bibnamefont
  {Jaatinen}}\ and\ \bibinfo {author} {\bibfnamefont {D.~J.}\ \bibnamefont
  {Hopper}},\ }\href {https://doi.org/10.1016/j.optlaseng.2007.06.011}
  {\bibfield  {journal} {\bibinfo  {journal} {Optics and Lasers in
  Engineering}\ }\textbf {\bibinfo {volume} {46}},\ \bibinfo {pages} {69}
  (\bibinfo {year} {2008})}\BibitemShut {NoStop}%
\bibitem [{\citenamefont {Negnevitsky}\ and\ \citenamefont
  {Turner}(2013)}]{negnevitsky_wideband_2013}%
  \BibitemOpen
  \bibfield  {author} {\bibinfo {author} {\bibfnamefont {V.}~\bibnamefont
  {Negnevitsky}}\ and\ \bibinfo {author} {\bibfnamefont {L.~D.}\ \bibnamefont
  {Turner}},\ }\href {https://doi.org/10.1364/OE.21.003103} {\bibfield
  {journal} {\bibinfo  {journal} {Optics Express}\ }\textbf {\bibinfo {volume}
  {21}},\ \bibinfo {pages} {3103} (\bibinfo {year} {2013})}\BibitemShut
  {NoStop}%
\bibitem [{\citenamefont {Wang}\ \emph {et~al.}(2022)\citenamefont {Wang},
  \citenamefont {Peng}, \citenamefont {Wang}, \citenamefont {Liu},\ and\
  \citenamefont {Guo}}]{Wang2022}%
  \BibitemOpen
  \bibfield  {author} {\bibinfo {author} {\bibfnamefont {B.}~\bibnamefont
  {Wang}}, \bibinfo {author} {\bibfnamefont {X.}~\bibnamefont {Peng}}, \bibinfo
  {author} {\bibfnamefont {H.}~\bibnamefont {Wang}}, \bibinfo {author}
  {\bibfnamefont {Y.}~\bibnamefont {Liu}},\ and\ \bibinfo {author}
  {\bibfnamefont {H.}~\bibnamefont {Guo}},\ }\href
  {https://doi.org/10.1063/5.0084605} {\bibfield  {journal} {\bibinfo
  {journal} {Review of Scientific Instruments}\ }\textbf {\bibinfo {volume}
  {93}},\ \bibinfo {pages} {043001} (\bibinfo {year} {2022})}\BibitemShut
  {NoStop}%
\bibitem [{\citenamefont {Sargsyan}\ \emph {et~al.}(2019)\citenamefont
  {Sargsyan}, \citenamefont {Amiryan}, \citenamefont {Pashayan-Leroy},
  \citenamefont {Leroy}, \citenamefont {Papoyan},\ and\ \citenamefont
  {Sarkisyan}}]{sargsyan_approach_2019}%
  \BibitemOpen
  \bibfield  {author} {\bibinfo {author} {\bibfnamefont {A.}~\bibnamefont
  {Sargsyan}}, \bibinfo {author} {\bibfnamefont {A.}~\bibnamefont {Amiryan}},
  \bibinfo {author} {\bibfnamefont {Y.}~\bibnamefont {Pashayan-Leroy}},
  \bibinfo {author} {\bibfnamefont {C.}~\bibnamefont {Leroy}}, \bibinfo
  {author} {\bibfnamefont {A.}~\bibnamefont {Papoyan}},\ and\ \bibinfo {author}
  {\bibfnamefont {D.}~\bibnamefont {Sarkisyan}},\ }\href
  {https://doi.org/10.1364/OL.44.005533} {\bibfield  {journal} {\bibinfo
  {journal} {Opt. Lett.}\ }\textbf {\bibinfo {volume} {44}},\ \bibinfo {pages}
  {5533} (\bibinfo {year} {2019})}\BibitemShut {NoStop}%
\bibitem [{\citenamefont {McGrew}\ \emph {et~al.}(2018)\citenamefont {McGrew},
  \citenamefont {Zhang}, \citenamefont {Fasano}, \citenamefont {Schäffer},
  \citenamefont {Beloy}, \citenamefont {Nicolodi}, \citenamefont {Brown},
  \citenamefont {Hinkley}, \citenamefont {Milani}, \citenamefont {Schioppo},
  \citenamefont {Yoon},\ and\ \citenamefont {Ludlow}}]{mcgrew_atomic_2018}%
  \BibitemOpen
  \bibfield  {author} {\bibinfo {author} {\bibfnamefont {W.~F.}\ \bibnamefont
  {McGrew}}, \bibinfo {author} {\bibfnamefont {X.}~\bibnamefont {Zhang}},
  \bibinfo {author} {\bibfnamefont {R.~J.}\ \bibnamefont {Fasano}}, \bibinfo
  {author} {\bibfnamefont {S.~A.}\ \bibnamefont {Schäffer}}, \bibinfo {author}
  {\bibfnamefont {K.}~\bibnamefont {Beloy}}, \bibinfo {author} {\bibfnamefont
  {D.}~\bibnamefont {Nicolodi}}, \bibinfo {author} {\bibfnamefont {R.~C.}\
  \bibnamefont {Brown}}, \bibinfo {author} {\bibfnamefont {N.}~\bibnamefont
  {Hinkley}}, \bibinfo {author} {\bibfnamefont {G.}~\bibnamefont {Milani}},
  \bibinfo {author} {\bibfnamefont {M.}~\bibnamefont {Schioppo}}, \bibinfo
  {author} {\bibfnamefont {T.~H.}\ \bibnamefont {Yoon}},\ and\ \bibinfo
  {author} {\bibfnamefont {A.~D.}\ \bibnamefont {Ludlow}},\ }\href
  {https://doi.org/10.1038/s41586-018-0738-2} {\bibfield  {journal} {\bibinfo
  {journal} {Nature}\ }\textbf {\bibinfo {volume} {564}},\ \bibinfo {pages}
  {87} (\bibinfo {year} {2018})}\BibitemShut {NoStop}%
\bibitem [{\citenamefont {Ma}\ \emph {et~al.}(2023)\citenamefont {Ma},
  \citenamefont {Qiao}, \citenamefont {Chen}, \citenamefont {Luo},
  \citenamefont {Yu}, \citenamefont {Wang}, \citenamefont {Lu}, \citenamefont
  {Zhao}, \citenamefont {Yang}, \citenamefont {Lin},\ and\ \citenamefont
  {Jiang}}]{ma-situ_2023}%
  \BibitemOpen
  \bibfield  {author} {\bibinfo {author} {\bibfnamefont {Y.}~\bibnamefont
  {Ma}}, \bibinfo {author} {\bibfnamefont {Z.}~\bibnamefont {Qiao}}, \bibinfo
  {author} {\bibfnamefont {Y.}~\bibnamefont {Chen}}, \bibinfo {author}
  {\bibfnamefont {G.}~\bibnamefont {Luo}}, \bibinfo {author} {\bibfnamefont
  {M.}~\bibnamefont {Yu}}, \bibinfo {author} {\bibfnamefont {Y.}~\bibnamefont
  {Wang}}, \bibinfo {author} {\bibfnamefont {D.}~\bibnamefont {Lu}}, \bibinfo
  {author} {\bibfnamefont {L.}~\bibnamefont {Zhao}}, \bibinfo {author}
  {\bibfnamefont {P.}~\bibnamefont {Yang}}, \bibinfo {author} {\bibfnamefont
  {Q.}~\bibnamefont {Lin}},\ and\ \bibinfo {author} {\bibfnamefont
  {Z.}~\bibnamefont {Jiang}},\ }\href {https://doi.org/10.1364/OE.483108}
  {\bibfield  {journal} {\bibinfo  {journal} {Opt. Express}\ }\textbf {\bibinfo
  {volume} {31}},\ \bibinfo {pages} {3743} (\bibinfo {year}
  {2023})}\BibitemShut {NoStop}%
\bibitem [{\citenamefont {Ding}\ \emph {et~al.}(2021)\citenamefont {Ding},
  \citenamefont {Shi}, \citenamefont {Zhang}, \citenamefont {Shen},
  \citenamefont {Zhang},\ and\ \citenamefont {Zhang}}]{PhysRevLett.126.083604}%
  \BibitemOpen
  \bibfield  {author} {\bibinfo {author} {\bibfnamefont {L.}~\bibnamefont
  {Ding}}, \bibinfo {author} {\bibfnamefont {K.}~\bibnamefont {Shi}}, \bibinfo
  {author} {\bibfnamefont {Q.}~\bibnamefont {Zhang}}, \bibinfo {author}
  {\bibfnamefont {D.}~\bibnamefont {Shen}}, \bibinfo {author} {\bibfnamefont
  {X.}~\bibnamefont {Zhang}},\ and\ \bibinfo {author} {\bibfnamefont
  {W.}~\bibnamefont {Zhang}},\ }\href
  {https://doi.org/10.1103/PhysRevLett.126.083604} {\bibfield  {journal}
  {\bibinfo  {journal} {Phys. Rev. Lett.}\ }\textbf {\bibinfo {volume} {126}},\
  \bibinfo {pages} {083604} (\bibinfo {year} {2021})}\BibitemShut {NoStop}%
\bibitem [{\citenamefont {Jaatinen}\ \emph {et~al.}(2009)\citenamefont
  {Jaatinen}, \citenamefont {Hopper},\ and\ \citenamefont
  {Back}}]{jaatinen_residual_2009}%
  \BibitemOpen
  \bibfield  {author} {\bibinfo {author} {\bibfnamefont {E.}~\bibnamefont
  {Jaatinen}}, \bibinfo {author} {\bibfnamefont {D.~J.}\ \bibnamefont
  {Hopper}},\ and\ \bibinfo {author} {\bibfnamefont {J.}~\bibnamefont {Back}},\
  }\bibfield  {journal} {\bibinfo  {journal} {Measurement Science and
  Technology}\ }\textbf {\bibinfo {volume} {20}},\ \href
  {https://doi.org/10.1088/0957-0233/20/2/025302}
  {10.1088/0957-0233/20/2/025302} (\bibinfo {year} {2009})\BibitemShut
  {NoStop}%
\bibitem [{\citenamefont {Sun}\ \emph {et~al.}(2016)\citenamefont {Sun},
  \citenamefont {Zhou}, \citenamefont {Zhou}, \citenamefont {Wang},\ and\
  \citenamefont {Zhan}}]{sun_modulation_2016}%
  \BibitemOpen
  \bibfield  {author} {\bibinfo {author} {\bibfnamefont {D.}~\bibnamefont
  {Sun}}, \bibinfo {author} {\bibfnamefont {C.}~\bibnamefont {Zhou}}, \bibinfo
  {author} {\bibfnamefont {L.}~\bibnamefont {Zhou}}, \bibinfo {author}
  {\bibfnamefont {J.}~\bibnamefont {Wang}},\ and\ \bibinfo {author}
  {\bibfnamefont {M.}~\bibnamefont {Zhan}},\ }\href
  {https://doi.org/10.1364/OE.24.010649} {\bibfield  {journal} {\bibinfo
  {journal} {Optics Express}\ }\textbf {\bibinfo {volume} {24}},\ \bibinfo
  {pages} {10649} (\bibinfo {year} {2016})}\BibitemShut {NoStop}%
\bibitem [{\citenamefont {Shen}\ \emph {et~al.}(2015)\citenamefont {Shen},
  \citenamefont {Li}, \citenamefont {Bi}, \citenamefont {Wang},\ and\
  \citenamefont {Chen}}]{shen_systematic_2015}%
  \BibitemOpen
  \bibfield  {author} {\bibinfo {author} {\bibfnamefont {H.}~\bibnamefont
  {Shen}}, \bibinfo {author} {\bibfnamefont {L.}~\bibnamefont {Li}}, \bibinfo
  {author} {\bibfnamefont {J.}~\bibnamefont {Bi}}, \bibinfo {author}
  {\bibfnamefont {J.}~\bibnamefont {Wang}},\ and\ \bibinfo {author}
  {\bibfnamefont {L.}~\bibnamefont {Chen}},\ }\href
  {https://doi.org/10.1103/PhysRevA.92.063809} {\bibfield  {journal} {\bibinfo
  {journal} {Physical Review A}\ }\textbf {\bibinfo {volume} {92}},\ \bibinfo
  {pages} {063809} (\bibinfo {year} {2015})}\BibitemShut {NoStop}%
\bibitem [{\citenamefont {Carminati}\ and\ \citenamefont
  {Scandurra}(2021)}]{carminati_impact_2021}%
  \BibitemOpen
  \bibfield  {author} {\bibinfo {author} {\bibfnamefont {M.}~\bibnamefont
  {Carminati}}\ and\ \bibinfo {author} {\bibfnamefont {G.}~\bibnamefont
  {Scandurra}},\ }\href {https://doi.org/10.1063/5.0050999} {\bibfield
  {journal} {\bibinfo  {journal} {Review of Scientific Instruments}\ }\textbf
  {\bibinfo {volume} {92}},\ \bibinfo {pages} {091501} (\bibinfo {year}
  {2021})}\BibitemShut {NoStop}%
\bibitem [{\citenamefont {Avalos}\ \emph {et~al.}(2023)\citenamefont {Avalos},
  \citenamefont {Nie}, \citenamefont {Yang}, \citenamefont {He}, \citenamefont
  {Kumar},\ and\ \citenamefont
  {Dieckmann}}]{avalos_field-programmable-gate-array-based_2023}%
  \BibitemOpen
  \bibfield  {author} {\bibinfo {author} {\bibfnamefont {V.}~\bibnamefont
  {Avalos}}, \bibinfo {author} {\bibfnamefont {X.}~\bibnamefont {Nie}},
  \bibinfo {author} {\bibfnamefont {A.}~\bibnamefont {Yang}}, \bibinfo {author}
  {\bibfnamefont {C.}~\bibnamefont {He}}, \bibinfo {author} {\bibfnamefont
  {S.}~\bibnamefont {Kumar}},\ and\ \bibinfo {author} {\bibfnamefont
  {K.}~\bibnamefont {Dieckmann}},\ }\href {https://doi.org/10.1063/5.0152305}
  {\bibfield  {journal} {\bibinfo  {journal} {Review of Scientific
  Instruments}\ }\textbf {\bibinfo {volume} {94}},\ \bibinfo {pages} {063001}
  (\bibinfo {year} {2023})}\BibitemShut {NoStop}%
\bibitem [{\citenamefont {Wu}\ \emph {et~al.}()\citenamefont {Wu},
  \citenamefont {Yan}, \citenamefont {Wei}, \citenamefont {Ma}, \citenamefont
  {Tu}, \citenamefont {Zhang},\ and\ \citenamefont
  {Wang}}]{wu_modulation_2018}%
  \BibitemOpen
  \bibfield  {author} {\bibinfo {author} {\bibfnamefont {C.~F.}\ \bibnamefont
  {Wu}}, \bibinfo {author} {\bibfnamefont {X.~S.}\ \bibnamefont {Yan}},
  \bibinfo {author} {\bibfnamefont {L.~X.}\ \bibnamefont {Wei}}, \bibinfo
  {author} {\bibfnamefont {P.}~\bibnamefont {Ma}}, \bibinfo {author}
  {\bibfnamefont {J.-H.}\ \bibnamefont {Tu}}, \bibinfo {author} {\bibfnamefont
  {J.-W.}\ \bibnamefont {Zhang}},\ and\ \bibinfo {author} {\bibfnamefont
  {L.-J.}\ \bibnamefont {Wang}},\ }\href
  {https://doi.org/10.1088/1674-1056/27/11/114203} {\bibfield  {journal}
  {\bibinfo  {journal} {Chinese Physics B}\ }\textbf {\bibinfo {volume} {27}},\
  \bibinfo {pages} {114203}}\BibitemShut {NoStop}%
\bibitem [{\citenamefont {Kimball}\ \emph {et~al.}(1999)\citenamefont
  {Kimball}, \citenamefont {Clyde}, \citenamefont {Budker}, \citenamefont
  {DeMille}, \citenamefont {Freedman}, \citenamefont {Rochester}, \citenamefont
  {Stalnaker},\ and\ \citenamefont {Zolotorev}}]{PhysRevA.60.1103}%
  \BibitemOpen
  \bibfield  {author} {\bibinfo {author} {\bibfnamefont {D.~F.}\ \bibnamefont
  {Kimball}}, \bibinfo {author} {\bibfnamefont {D.}~\bibnamefont {Clyde}},
  \bibinfo {author} {\bibfnamefont {D.}~\bibnamefont {Budker}}, \bibinfo
  {author} {\bibfnamefont {D.}~\bibnamefont {DeMille}}, \bibinfo {author}
  {\bibfnamefont {S.~J.}\ \bibnamefont {Freedman}}, \bibinfo {author}
  {\bibfnamefont {S.}~\bibnamefont {Rochester}}, \bibinfo {author}
  {\bibfnamefont {J.~E.}\ \bibnamefont {Stalnaker}},\ and\ \bibinfo {author}
  {\bibfnamefont {M.}~\bibnamefont {Zolotorev}},\ }\href
  {https://doi.org/10.1103/PhysRevA.60.1103} {\bibfield  {journal} {\bibinfo
  {journal} {Phys. Rev. A}\ }\textbf {\bibinfo {volume} {60}},\ \bibinfo
  {pages} {1103} (\bibinfo {year} {1999})}\BibitemShut {NoStop}%
\bibitem [{\citenamefont {Zhou}\ \emph {et~al.}(2020)\citenamefont {Zhou},
  \citenamefont {Zhang}, \citenamefont {Luo},\ and\ \citenamefont
  {Xu}}]{PhysRevA.101.062506}%
  \BibitemOpen
  \bibfield  {author} {\bibinfo {author} {\bibfnamefont {M.}~\bibnamefont
  {Zhou}}, \bibinfo {author} {\bibfnamefont {S.}~\bibnamefont {Zhang}},
  \bibinfo {author} {\bibfnamefont {L.}~\bibnamefont {Luo}},\ and\ \bibinfo
  {author} {\bibfnamefont {X.}~\bibnamefont {Xu}},\ }\href
  {https://doi.org/10.1103/PhysRevA.101.062506} {\bibfield  {journal} {\bibinfo
   {journal} {Phys. Rev. A}\ }\textbf {\bibinfo {volume} {101}},\ \bibinfo
  {pages} {062506} (\bibinfo {year} {2020})}\BibitemShut {NoStop}%
\bibitem [{\citenamefont {Jang}\ \emph {et~al.}(2014)\citenamefont {Jang},
  \citenamefont {Na}, \citenamefont {Moon},\ and\ \citenamefont
  {Yoon}}]{PhysRevA.89.062510}%
  \BibitemOpen
  \bibfield  {author} {\bibinfo {author} {\bibfnamefont {G.~H.}\ \bibnamefont
  {Jang}}, \bibinfo {author} {\bibfnamefont {M.}~\bibnamefont {Na}}, \bibinfo
  {author} {\bibfnamefont {B.}~\bibnamefont {Moon}},\ and\ \bibinfo {author}
  {\bibfnamefont {T.~H.}\ \bibnamefont {Yoon}},\ }\href
  {https://doi.org/10.1103/PhysRevA.89.062510} {\bibfield  {journal} {\bibinfo
  {journal} {Phys. Rev. A}\ }\textbf {\bibinfo {volume} {89}},\ \bibinfo
  {pages} {062510} (\bibinfo {year} {2014})}\BibitemShut {NoStop}%
\bibitem [{\citenamefont {Qiao}\ \emph {et~al.}(2023)\citenamefont {Qiao},
  \citenamefont {Liu}, \citenamefont {Zhou}, \citenamefont {Luo},\ and\
  \citenamefont {Xu}}]{qiao_investigation_2023}%
  \BibitemOpen
  \bibfield  {author} {\bibinfo {author} {\bibfnamefont {H.}~\bibnamefont
  {Qiao}}, \bibinfo {author} {\bibfnamefont {L.}~\bibnamefont {Liu}}, \bibinfo
  {author} {\bibfnamefont {M.}~\bibnamefont {Zhou}}, \bibinfo {author}
  {\bibfnamefont {L.}~\bibnamefont {Luo}},\ and\ \bibinfo {author}
  {\bibfnamefont {X.}~\bibnamefont {Xu}},\ }\href
  {https://doi.org/10.1063/5.0155776} {\bibfield  {journal} {\bibinfo
  {journal} {Applied Physics Letters}\ }\textbf {\bibinfo {volume} {122}},\
  \bibinfo {pages} {224002} (\bibinfo {year} {2023})}\BibitemShut {NoStop}%
\bibitem [{\citenamefont {Ejtemaee}\ \emph {et~al.}(2010)\citenamefont
  {Ejtemaee}, \citenamefont {Thomas},\ and\ \citenamefont
  {Haljan}}]{PhysRevA.82.063419}%
  \BibitemOpen
  \bibfield  {author} {\bibinfo {author} {\bibfnamefont {S.}~\bibnamefont
  {Ejtemaee}}, \bibinfo {author} {\bibfnamefont {R.}~\bibnamefont {Thomas}},\
  and\ \bibinfo {author} {\bibfnamefont {P.~C.}\ \bibnamefont {Haljan}},\
  }\href {https://doi.org/10.1103/PhysRevA.82.063419} {\bibfield  {journal}
  {\bibinfo  {journal} {Phys. Rev. A}\ }\textbf {\bibinfo {volume} {82}},\
  \bibinfo {pages} {063419} (\bibinfo {year} {2010})}\BibitemShut {NoStop}%
\bibitem [{\citenamefont {Gulde}(2003)}]{blade1}%
  \BibitemOpen
  \bibfield  {author} {\bibinfo {author} {\bibfnamefont {S.}~\bibnamefont
  {Gulde}},\ }\href@noop {} {Ph.D. thesis},\ \bibinfo  {school} {Universität
  Innsbruck} (\bibinfo {year} {2003})\BibitemShut {NoStop}%
\bibitem [{\citenamefont {Mizrahi}(2013)}]{blade2}%
  \BibitemOpen
  \bibfield  {author} {\bibinfo {author} {\bibfnamefont {J.~A.}\ \bibnamefont
  {Mizrahi}},\ }\href@noop {} {Ph.D. thesis},\ \bibinfo  {school} {University
  of Maryland} (\bibinfo {year} {2013})\BibitemShut {NoStop}%
\bibitem [{\citenamefont {Yao}(2009)}]{4785283}%
  \BibitemOpen
  \bibfield  {author} {\bibinfo {author} {\bibfnamefont {J.}~\bibnamefont
  {Yao}},\ }\href {https://doi.org/10.1109/JLT.2008.2009551} {\bibfield
  {journal} {\bibinfo  {journal} {Journal of Lightwave Technology}\ }\textbf
  {\bibinfo {volume} {27}},\ \bibinfo {pages} {314} (\bibinfo {year}
  {2009})}\BibitemShut {NoStop}%
\bibitem [{\citenamefont {Uehara}\ \emph {et~al.}(2014)\citenamefont {Uehara},
  \citenamefont {Tsuji}, \citenamefont {Hagiwara},\ and\ \citenamefont
  {Onodera}}]{Uehara2014OpticalBF}%
  \BibitemOpen
  \bibfield  {author} {\bibinfo {author} {\bibfnamefont {T.}~\bibnamefont
  {Uehara}}, \bibinfo {author} {\bibfnamefont {K.}~\bibnamefont {Tsuji}},
  \bibinfo {author} {\bibfnamefont {K.}~\bibnamefont {Hagiwara}},\ and\
  \bibinfo {author} {\bibfnamefont {N.}~\bibnamefont {Onodera}},\ }\href
  {https://api.semanticscholar.org/CorpusID:42272509} {\bibfield  {journal}
  {\bibinfo  {journal} {Optical Engineering}\ }\textbf {\bibinfo {volume} {53}}
  (\bibinfo {year} {2014})}\BibitemShut {NoStop}%
\bibitem [{\citenamefont {Shen}\ \emph {et~al.}(2020)\citenamefont {Shen},
  \citenamefont {Ding}, \citenamefont {Zhang}, \citenamefont {Zhu},
  \citenamefont {Wang}, \citenamefont {Zhang},\ and\ \citenamefont
  {Zhang}}]{Shen_2020}%
  \BibitemOpen
  \bibfield  {author} {\bibinfo {author} {\bibfnamefont {D.}~\bibnamefont
  {Shen}}, \bibinfo {author} {\bibfnamefont {L.}~\bibnamefont {Ding}}, \bibinfo
  {author} {\bibfnamefont {Q.}~\bibnamefont {Zhang}}, \bibinfo {author}
  {\bibfnamefont {C.}~\bibnamefont {Zhu}}, \bibinfo {author} {\bibfnamefont
  {Y.}~\bibnamefont {Wang}}, \bibinfo {author} {\bibfnamefont {W.}~\bibnamefont
  {Zhang}},\ and\ \bibinfo {author} {\bibfnamefont {X.}~\bibnamefont {Zhang}},\
  }\href {https://doi.org/10.1088/1674-1056/ab8c41} {\bibfield  {journal}
  {\bibinfo  {journal} {Chinese Physics B}\ }\textbf {\bibinfo {volume} {29}},\
  \bibinfo {pages} {074210} (\bibinfo {year} {2020})}\BibitemShut {NoStop}%
\bibitem [{\citenamefont {Saleh}\ \emph {et~al.}(2015)\citenamefont {Saleh},
  \citenamefont {Millo}, \citenamefont {Didier}, \citenamefont {Kersal\'{e}},\
  and\ \citenamefont {Lacro\^{u}te}}]{Saleh:15}%
  \BibitemOpen
  \bibfield  {author} {\bibinfo {author} {\bibfnamefont {K.}~\bibnamefont
  {Saleh}}, \bibinfo {author} {\bibfnamefont {J.}~\bibnamefont {Millo}},
  \bibinfo {author} {\bibfnamefont {A.}~\bibnamefont {Didier}}, \bibinfo
  {author} {\bibfnamefont {Y.}~\bibnamefont {Kersal\'{e}}},\ and\ \bibinfo
  {author} {\bibfnamefont {C.}~\bibnamefont {Lacro\^{u}te}},\ }\href
  {https://doi.org/10.1364/AO.54.009446} {\bibfield  {journal} {\bibinfo
  {journal} {Appl. Opt.}\ }\textbf {\bibinfo {volume} {54}},\ \bibinfo {pages}
  {9446} (\bibinfo {year} {2015})}\BibitemShut {NoStop}%
\bibitem [{\citenamefont {Couturier}\ \emph {et~al.}(2018)\citenamefont
  {Couturier}, \citenamefont {Nosske}, \citenamefont {Hu}, \citenamefont {Tan},
  \citenamefont {Qiao}, \citenamefont {Jiang}, \citenamefont {Chen},\ and\
  \citenamefont {Weidemüller}}]{10.1063/1.5025537}%
  \BibitemOpen
  \bibfield  {author} {\bibinfo {author} {\bibfnamefont {L.}~\bibnamefont
  {Couturier}}, \bibinfo {author} {\bibfnamefont {I.}~\bibnamefont {Nosske}},
  \bibinfo {author} {\bibfnamefont {F.}~\bibnamefont {Hu}}, \bibinfo {author}
  {\bibfnamefont {C.}~\bibnamefont {Tan}}, \bibinfo {author} {\bibfnamefont
  {C.}~\bibnamefont {Qiao}}, \bibinfo {author} {\bibfnamefont {Y.~H.}\
  \bibnamefont {Jiang}}, \bibinfo {author} {\bibfnamefont {P.}~\bibnamefont
  {Chen}},\ and\ \bibinfo {author} {\bibfnamefont {M.}~\bibnamefont
  {Weidemüller}},\ }\href {https://doi.org/10.1063/1.5025537} {\bibfield
  {journal} {\bibinfo  {journal} {Review of Scientific Instruments}\ }\textbf
  {\bibinfo {volume} {89}},\ \bibinfo {pages} {043103} (\bibinfo {year}
  {2018})}\BibitemShut {NoStop}%
\bibitem [{\citenamefont {Moon}\ and\ \citenamefont {Noh}(2018)}]{Moon:18}%
  \BibitemOpen
  \bibfield  {author} {\bibinfo {author} {\bibfnamefont {G.}~\bibnamefont
  {Moon}}\ and\ \bibinfo {author} {\bibfnamefont {H.-R.}\ \bibnamefont {Noh}},\
  }\href {https://doi.org/10.1364/AO.57.003881} {\bibfield  {journal} {\bibinfo
   {journal} {Appl. Opt.}\ }\textbf {\bibinfo {volume} {57}},\ \bibinfo {pages}
  {3881} (\bibinfo {year} {2018})}\BibitemShut {NoStop}%
\bibitem [{\citenamefont {Noh}(2016)}]{Heung-Ryoul}%
  \BibitemOpen
  \bibfield  {author} {\bibinfo {author} {\bibfnamefont {H.-R.}\ \bibnamefont
  {Noh}},\ }\href {https://doi.org/10.1364/JOSAB.33.000308} {\bibfield
  {journal} {\bibinfo  {journal} {Journal of the Optical Society of America B}\
  }\textbf {\bibinfo {volume} {33}},\ \bibinfo {pages} {308} (\bibinfo {year}
  {2016})}\BibitemShut {NoStop}%
\bibitem [{\citenamefont {Jackson}\ \emph {et~al.}(2018)\citenamefont
  {Jackson}, \citenamefont {Sawaoka}, \citenamefont {Bhatt}, \citenamefont
  {Potnis},\ and\ \citenamefont {Vutha}}]{jackson_laser_2018}%
  \BibitemOpen
  \bibfield  {author} {\bibinfo {author} {\bibfnamefont {S.}~\bibnamefont
  {Jackson}}, \bibinfo {author} {\bibfnamefont {H.}~\bibnamefont {Sawaoka}},
  \bibinfo {author} {\bibfnamefont {N.}~\bibnamefont {Bhatt}}, \bibinfo
  {author} {\bibfnamefont {S.}~\bibnamefont {Potnis}},\ and\ \bibinfo {author}
  {\bibfnamefont {A.~C.}\ \bibnamefont {Vutha}},\ }\href
  {https://doi.org/10.1063/1.5012000} {\bibfield  {journal} {\bibinfo
  {journal} {Review of Scientific Instruments}\ }\textbf {\bibinfo {volume}
  {89}},\ \bibinfo {pages} {033109} (\bibinfo {year} {2018})}\BibitemShut
  {NoStop}%
\bibitem [{\citenamefont {Subhankar}\ \emph {et~al.}(2019)\citenamefont
  {Subhankar}, \citenamefont {Restelli}, \citenamefont {Wang}, \citenamefont
  {Rolston},\ and\ \citenamefont {Porto}}]{subhankar_microcontroller_2019}%
  \BibitemOpen
  \bibfield  {author} {\bibinfo {author} {\bibfnamefont {S.}~\bibnamefont
  {Subhankar}}, \bibinfo {author} {\bibfnamefont {A.}~\bibnamefont {Restelli}},
  \bibinfo {author} {\bibfnamefont {Y.}~\bibnamefont {Wang}}, \bibinfo {author}
  {\bibfnamefont {S.~L.}\ \bibnamefont {Rolston}},\ and\ \bibinfo {author}
  {\bibfnamefont {J.~V.}\ \bibnamefont {Porto}},\ }\href
  {https://doi.org/10.1063/1.5067266} {\bibfield  {journal} {\bibinfo
  {journal} {Review of Scientific Instruments}\ }\textbf {\bibinfo {volume}
  {90}},\ \bibinfo {pages} {043115} (\bibinfo {year} {2019})}\BibitemShut
  {NoStop}%
\bibitem [{\citenamefont {Zeng}\ \emph {et~al.}(2021)\citenamefont {Zeng},
  \citenamefont {Fu}, \citenamefont {Liu}, \citenamefont {He}, \citenamefont
  {Liu}, \citenamefont {Xu}, \citenamefont {Sun},\ and\ \citenamefont
  {Wang}}]{zeng_stabilizing_2021}%
  \BibitemOpen
  \bibfield  {author} {\bibinfo {author} {\bibfnamefont {Y.}~\bibnamefont
  {Zeng}}, \bibinfo {author} {\bibfnamefont {Z.}~\bibnamefont {Fu}}, \bibinfo
  {author} {\bibfnamefont {Y.-Y.}\ \bibnamefont {Liu}}, \bibinfo {author}
  {\bibfnamefont {X.-D.}\ \bibnamefont {He}}, \bibinfo {author} {\bibfnamefont
  {M.}~\bibnamefont {Liu}}, \bibinfo {author} {\bibfnamefont {P.}~\bibnamefont
  {Xu}}, \bibinfo {author} {\bibfnamefont {X.-H.}\ \bibnamefont {Sun}},\ and\
  \bibinfo {author} {\bibfnamefont {J.}~\bibnamefont {Wang}},\ }\href
  {https://doi.org/10.1364/AO.415011} {\bibfield  {journal} {\bibinfo
  {journal} {Applied Optics}\ }\textbf {\bibinfo {volume} {60}},\ \bibinfo
  {pages} {1159} (\bibinfo {year} {2021})}\BibitemShut {NoStop}%
\bibitem [{\citenamefont {Wang}\ \emph {et~al.}(2020)\citenamefont {Wang},
  \citenamefont {Feng}, \citenamefont {Xu},\ and\ \citenamefont
  {Ni}}]{wang_steady-state_2020}%
  \BibitemOpen
  \bibfield  {author} {\bibinfo {author} {\bibfnamefont {X.}~\bibnamefont
  {Wang}}, \bibinfo {author} {\bibfnamefont {L.}~\bibnamefont {Feng}}, \bibinfo
  {author} {\bibfnamefont {J.}~\bibnamefont {Xu}},\ and\ \bibinfo {author}
  {\bibfnamefont {P.}~\bibnamefont {Ni}},\ }\href
  {https://doi.org/10.1364/AO.379557} {\bibfield  {journal} {\bibinfo
  {journal} {Applied Optics}\ }\textbf {\bibinfo {volume} {59}},\ \bibinfo
  {pages} {1347} (\bibinfo {year} {2020})}\BibitemShut {NoStop}%
\bibitem [{\citenamefont {Bell}\ \emph {et~al.}(1991)\citenamefont {Bell},
  \citenamefont {Gill}, \citenamefont {Klein}, \citenamefont {Levick},
  \citenamefont {Tamm},\ and\ \citenamefont {Schnier}}]{PhysRevA.44.R20}%
  \BibitemOpen
  \bibfield  {author} {\bibinfo {author} {\bibfnamefont {A.~S.}\ \bibnamefont
  {Bell}}, \bibinfo {author} {\bibfnamefont {P.}~\bibnamefont {Gill}}, \bibinfo
  {author} {\bibfnamefont {H.~A.}\ \bibnamefont {Klein}}, \bibinfo {author}
  {\bibfnamefont {A.~P.}\ \bibnamefont {Levick}}, \bibinfo {author}
  {\bibfnamefont {C.}~\bibnamefont {Tamm}},\ and\ \bibinfo {author}
  {\bibfnamefont {D.}~\bibnamefont {Schnier}},\ }\href
  {https://doi.org/10.1103/PhysRevA.44.R20} {\bibfield  {journal} {\bibinfo
  {journal} {Phys. Rev. A}\ }\textbf {\bibinfo {volume} {44}},\ \bibinfo
  {pages} {R20} (\bibinfo {year} {1991})}\BibitemShut {NoStop}%
\bibitem [{\citenamefont {Metcalf}\ and\ \citenamefont {Van
  Der~Straten}(1999)}]{metcalf_laser_1999}%
  \BibitemOpen
  \bibfield  {author} {\bibinfo {author} {\bibfnamefont {H.~J.}\ \bibnamefont
  {Metcalf}}\ and\ \bibinfo {author} {\bibfnamefont {P.}~\bibnamefont {Van
  Der~Straten}},\ }\href {https://doi.org/10.1007/978-1-4612-1470-0} {\emph
  {\bibinfo {title} {Laser {Cooling} and {Trapping}}}}\ (\bibinfo  {publisher}
  {Springer New York},\ \bibinfo {address} {New York, NY},\ \bibinfo {year}
  {1999})\BibitemShut {NoStop}%
\end{thebibliography}%

\end{document}